\documentclass[aps,prl,10pt,twocolumn,showpacs,superscriptaddress]{revtex4-1}
\usepackage{amsmath}
\usepackage{latexsym}
\usepackage{amssymb}
\usepackage{bm}
\usepackage{graphics,epstopdf}
\usepackage{color}
\usepackage{hyperref}

\usepackage{newlfont}
\usepackage{amsfonts}
\usepackage{amsthm}
\usepackage{graphicx}
\usepackage{epsfig}

\newcommand{\ket}[1]{|{#1}\rangle}
\newcommand{\bra}[1]{\langle{#1}|}

\newcommand{\expec}[1]{\langle#1\rangle}

\usepackage{times}

\usepackage[up]{subfigure}

\newcommand{\be}{\begin{equation}}
\newcommand{\ee}{\end{equation}}
\newcommand{\bc}{\begin{center}}
\newcommand{\ec}{\end{center}}
\newcommand{\bea}{\begin{eqnarray}}
\newcommand{\eea}{\end{eqnarray}}
\newcommand{\ba}{\begin{array}}
\newcommand{\ea}{\end{array}}

\begin{document}
\title{Alteration in Non-Classicality of Light on Passing Through a Linear Polarization Beam Splitter}

\author{Namrata Shukla}
\email{namratashukla@hri.res.in}

\affiliation{Now at, Quantum Information and Computation Group,\\
Harish-Chandra Research Institute, Chhatnag Road, Jhunsi,
Allahabad 211 019, India}

\affiliation{Department of Physics,\\
University of Allahabad, Allahabad, Allahabad-211001, UP, India}

%\author{Lorenzo Maccone}
%\affiliation{ Dip.~Fisica and INFN Sez.~Pavia, \\
%University ~of Pavia, via Bassi 6, I-27100 Pavia, Italy }

%\author{Uttam Singh}
%\email{uttamsingh@hri.res.in}
%\affiliation{Quantum Information and Computation Group,\\
%Harish-Chandra Research Institute, Chhatnag Road, Jhunsi,
%Allahabad 211 019, India}

\author{Ranjana Prakash}
%\email{akpati@hri.res.in}

%\author{Uttam Singh}
%\email{uttamsingh@hri.res.in}

\affiliation{Department of Physics,\\
University of Allahabad, Allahabad, Allahabad-211001, UP, India}

%\author{}
%\email{}

%\affiliation{}

\date{\today}

\begin{abstract}
We observe the polarization squeezing in the mixture of a two mode squeezed vacuum and a simple 
coherent light through a linear polarization beam splitter.
Squeezed vacuum not being squeezed in polarization, 
generates polarization squeezed light when superposed with coherent light. All the three Stokes 
parameters of the light produced on the output port of polarization beam splitter are found to 
be squeezed and squeezing factor also depends upon the parameters of coherent light.
\end{abstract}

\maketitle 

\section{Introduction}
In classical optics, the state of polarization of a light beam is visualized by Stokes vector on the
Poincare sphere \cite{stokes,Born-wolf} and it is determined by four Stokes parameters $S_0$ and 
${\bm S}=(S_1, S_2, S_3)$, following the relation ${\bm S}^2={S_1}^2+{S_2}^2+{S_3}^2 $. These Stokes parameters 
involve coherence functions \cite{mandel-wolf} of order (1,1) and it has been realized that these are insufficient
to describe polarization completely as $\bm{S}=0$ does not represent only 
unpolarized light \cite{prakash-chandra}. But, these parameters still remain important because of their role in 
non-classicalities associated with polarization. Quantum mechanical analogue of Stokes parameters can also be defined to characterize quantum nature 
of polarization. These quantum Stokes operators hermitian in nature and act as observables for the system. 
These hermitian Stokes operators can be defined as the quantum versions of their classical counterparts and these are given by
\begin{equation}
\label{eq1}
\hat S_{0, 1}=\hat a_{x}^\dagger \hat a_{x}\pm \hat a_{y}^\dagger \hat a_{y},~
\hat S_{2}+i \hat S_{3}=2 \hat a_{x}^\dagger \hat a_{y},
\end {equation}
where $ \hat a_{x,y}, \hat a_{x,y}^\dagger $ refer to the photon annihilation and creation operators 
respectively of the two orthogonal polarization modes $x$ and $y$ satisfying the commutation relations
$ [\hat a_{j}, \hat a_{k}^\dagger]=\delta_{jk} $ for $j,k=x,y$. 
The mean value of the radius of quantum Poincare sphere is given by square
root of expectation value of either side of the equation
\begin{equation}
\label{eq2}
\hat S_{1}^2+\hat S_{2}^2+\hat S_{3}^2=\hat S_{0}^2+2\hat S_0.
\end{equation}
 Note that, Stokes operators do not commute with each other and hence this equation has an extra term 
 $2\hat S_0$ on the right hand side. Thus, a quantum polarization state is described by 
the four Stokes operators, $\hat S_{0}$ which is the total photon number operator denoting the beam intensity
and $\hat{\bm S}=\hat S_{1}, \hat S_{2}, \hat S_{3}$ with commutation relations
\begin{equation}
\label{eq3}
[\hat S_0, \hat S_j]=0, [\hat S_j, \hat S_k]=2i{\sum_{l}\epsilon_{jkl}}~\hat S_{l},
\end {equation}                                                                                            
following the SU(2) algebra. Here, $\epsilon_{jkl}$ is Levi-Civita symbol for $(j,k,l=1,2,$ or $3,~j\neq k\neq l\neq j)$.
These relations are parallel to the commutation relations for components of the angular momentum 
operator. These non-zero commutators show that simultaneous exact measurement of the quantities  
represented by these Stokes operators are impossible and the following uncertainty relations hold
\begin{equation}
\label{eq4}
V_{j} V_{k}\geqslant{\expec{\hat S_{l}}}^2,~ 
V_{j}\equiv\expec{\hat S_{j}^2}-{\expec{\hat S_{j}}}^2.
\end{equation}
Here $V_{j}$ stands for the variance $ \expec{\hat S_{j}^2}- {\expec{\hat S_{j}}}^2 $ of the quantum 
Stokes operators $\hat S_{j}$. \\
 
 Polarization squeezing is a non-classicality similar to ordinary squeezing, and is defined 
 almost in a similar fashion using Stokes operators as the continuous variables for the system 
 describing polarization. Polarization squeezing first introduced by Chirkin et al. \cite{chirkin} was initially defined 
 using the commutation relations followed by the Stokes operators. Later, this definition was modified by 
 Heersink et al. \cite{heersink} taking into account the uncertainty relations followed by these Stokes operators 
 and generalized by Luis and Korolkova \cite{luis-korolkova} for a general component of Stokes operator vector. 
 The authors have written the criterion for polarization squeezing for a general component of Stokes vector 
 operator $\hat S_{\bm n}$ along the unit vector $ \bm n$ \cite{prakash-shukla,prakash-shukla1,prakash-shukla2} 
 in the following form 
\begin{eqnarray}
\label{eq5}
V_{\bm n}\equiv\expec{\Delta \hat S_{\bm n}^2} &<&{|\expec{\hat S_{\bm n_{\perp}}}|}_{max}\nonumber\\
&=&\sqrt{{|\expec{\hat {\bm S}}|}^2-{\expec{S_{\bm n}}}^2}
\end{eqnarray}
arguing that, there are infinite directions $ \bm n_{\perp} $ for a given component $ \hat S_{\bm n} $ 
and therefore one needs to consider the maximum possible value of $ |\expec{\hat S_{\bm n_{\perp}}}|$.  
All of the above definitions have been used in different studies on polarization 
squeezing \cite{polsq1,polsq2,polsq3} in the order of improvement considering 
the uncertainty relations and characterization of polarization squeezing in a general component of Stokes operator vector. 
In our study, we use the criterion in Eq.~(\ref{eq5}) for polarization squeezing, which is the most general and 
based on the actual uncertainty relations. Squeezing factor $ \mathcal{S}_{\bm n}$ and 
degree of squeezing $ \mathcal{D}_{\bm n} $ to measure polarization squeezing can be defined as
\begin{equation}
\label{eq6}
\mathcal{S}_{\bm n}=\frac{V_{\bm n}}{\sqrt{{|\expec{\hat {\bm S}}|}^2-{\expec{\hat S_{\bm n}}}^2}},~
\mathcal{D}_{\bm n}=1-\mathcal{S}_{\bm n},
\end{equation}
respectively.
Non-classicalities are seen when $ 1>\mathcal{S}_{\bm n}>0 $ and the degree of squeezing 
$ \mathcal{D}_{\bm n}$ lies between $0$ and $1$. \\

Since, quantum Stokes operators and non-classical polarization can be used for quantum 
information protocols in quantum communication, 
it is important to study polarization squeezing in such systems. 
Another convenience reason being the easy measurement of 
Stokes parameters using linear optical elements, polarization squeezing is easy to experimentally measure. 
The direct measurement schemes are developed methods 
for measuring these parameters and they preserve quantum noise property.\\

\begin{figure}[bt]
\centering
\includegraphics[width=\linewidth]{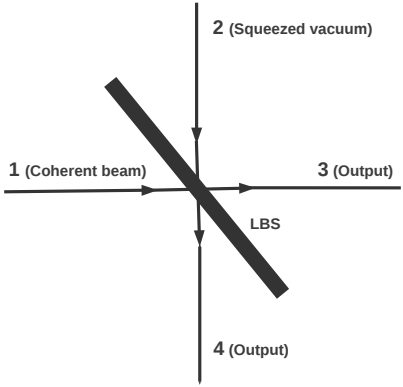}
\vspace{-5cm}
\caption{Superposition of coherent light beam and squeezed vaccum through linear polarization beam splitter.}
\label{f1}
\end{figure}

In the present paper, we study a process where linear beam splitter mixes coherent light with a 
two mode squeezed vacuum and it is observed that, the output beam from the beam splitter exhibits polarization
squeezing. Illustration in Fig.~\ref{f1} shows the superposition on linear polarization beam splitter.
The input non-classical light is a two mode squeezed vacuum which does not show polarization
squeezing. The linear beam splitter can not convert classical light beams into non-classical
though nonlinear beam splitters do. The polarization squeezing at the output port $\bm 3$ therefore shows
that, the input non-classical beam gives non-classicality in the output. \\

If the two mode coherent light beam incident at ports $\bm 1$ and two mode squeezed vacuum at port $\bm2$, 
are represented by annihilation operators $\hat a_x, \hat a_y$ and $\hat b_x, \hat b_y $, respectively and output at ports $\bm 3$ and 
$\bm 4$ have annihilation operators $\hat c_{x,y}$ and $\hat d_{x,y}$, respectively, for the two mode coherent state $\ket{\alpha_x,\alpha_y}$ we have
\begin{equation}
\label{eq7}
\hat a_x\ket{\alpha_x,\alpha_y}=\alpha_x\ket{\alpha_x,\alpha_y}, ~
\hat a_y\ket{\alpha_x,\alpha_y}=\alpha_y\ket{\alpha_x,\alpha_y},
\end{equation}
and for two mode squeezed vacuum
\begin{equation}
\label{eq8}
 \hat b_x(t)=c~\hat b_x\ket{0}+is~b_{y}^\dagger\ket{0},~ \hat b_y(t)=c~\hat b_y\ket{0}+is~b_{x}^\dagger\ket{0},
\end{equation}
where $ c=\cosh kt,~s=\sinh kt $, $kt$ being the interaction time for nonlinear interaction. 
After this superposition, the beams at ports $\bm 3$ and $\bm 4$ can be represented as
\begin{equation}
\label{eqextra9}
\hat c=t\hat a+ir\hat b, ~\hat d=t\hat a+ir \hat a,
\end{equation}
where $t$ and $r$ are  the transmission coefficient and reflection coefficients, respectively. \\

In this problem, we are only interested in port $\bm 3$ and we can write the $x$ and $y$ modes of the 
annihilation operator at the output port $\bm3$ as
\begin{equation}
\label{eq9}
\hat c_x=t_x\hat a_x+ir_x\hat b_x, \hat c_y=t_y\hat a_y+ir_y\hat b_y,
\end{equation}
with $t_{x,y}$ and $r_{x,y}$ being the transmission coefficient and reflection coefficient, respectively for the two modes. \\

To have an idea about the non-classicality in mixing of two light beams, let us consider, the two input beams 
having density operators
\begin{equation}
\label{eq10}
\hat\rho_1=\int d^2\alpha~P_1(\alpha)\ket{\alpha}\bra{\alpha},~
\hat\rho_2=\int d^2\beta~P_2(\beta)\ket{\beta}\bra{\beta}.
\end{equation}
The composite density operator cab therefore be written as
\begin{eqnarray}
\hat\rho&=&\int d^2\alpha~d^2\beta~P_1(\alpha) P_2(\beta)\ket{\alpha,\beta}\bra{\alpha,\beta}\nonumber\\
&=&\int d^2\alpha~d^2\beta~P_1(\alpha)P_2(\beta)~\exp{[-(|\alpha|^2+|\beta|^2)]}\nonumber\\
&&\exp{[\alpha a^\dagger+\beta b^\dagger-{h.c.}]}\ket{0.0}\bra{0,0}\exp{[\alpha^* a +\beta^* b-{h.c.}]},\nonumber\\  
\label{eq11}
\end{eqnarray}
where {\it h.c.}stands for hermitian conjugate. 
This leads to
\begin{equation}
\label{eq12}
 \alpha\hat a^\dagger+\beta\hat b^\dagger=(t\alpha-ir\beta)\hat c^\dagger+(t\beta-ir\alpha)\hat d^\dagger,
\end{equation}
and hence
\begin{eqnarray}
\label{eq13}
\hat\rho&=&\int d^2\alpha~d^2\beta~P_1(\alpha) P_2(\beta)\ket{t\alpha-ir\beta,t\beta-ir\alpha}\nonumber\\
&&\bra{t\alpha-ir\beta,t\beta-ir\alpha} \nonumber\\
&=&\int d^2r~d^2\delta~P(r,\delta) \ket{r,\delta}\bra{r,\delta},
\end{eqnarray}
where $P(R,\delta)=P_1(tr+ir\delta)~P_2(t\delta+ir\delta)$.\\

This shows that, if $P_1$ and $P_2$ are non negative, $P$ also has a non-negative value and 
classical input light beams mix at a linear beam splitter to generate  classical output light beams. 
However, if one of $P_1$ and $P_2$ is not non-negative, $P$ would also be non-negative and 
input non-classical beam gives non-classicality in the output. Generation of anti-bunched light by 
mixing of squeezed light with classical light is a very well known example. We observe here the 
non-classicality in the form of polarization squeezing exhibited here in a similar manner. 

\section{The Two mode Squeezed vacuum and Polarization squeezing}
The two mode squeezed vacuum as shown in the Eq.~(\ref{eq8}) can also be represented as,
\begin{equation}
\label{eq14}
\hat b_x(t)=c \hat b_{x0}\ket{0}+isb_{y0}^\dagger\ket{0},~
\hat b_y(t)=c \hat b_{y0}\ket{0}+isb_{x0}^\dagger\ket{0}, 
\end{equation}
where $b_{x0}, b_{y0}$ are the annihilation operators initially and $ b_x(t), b_y(t)$ at time $t$ after the 
non-degenerate parametric amplification. 
For initial vacuum state, straight calculations give
\begin{eqnarray}
\label{eq15}
&& \expec{\hat S_0}=2s^2,\expec{\hat S_1}=\expec{\hat S_2}=\expec{\hat S_3}=0,\nonumber\\
&& \expec{{\hat S_1}^2}=\expec{{\hat S_2}^2}=\expec{{\hat S_3}^2}=4s^4+4s^2.
\end{eqnarray}
It is easy to see by plugging in the values in the inequality criterion for polarization squeezing given by Eq.~(\ref{eq5}) that, none of the 
Stokes operators show polarization squeezing. But, it is very important to notice that the squeezed vacuum is observed 
squeezed in polarization\cite{Korolkova} when the first criterion for polarization squeezing \cite{chirkin} is used, however,
the two mode squeezed vacuum shows no squeezing in polarization on using the general criterion that considers uncertainty relation 
followed by Stokes operators in addition to the commutation relations. We observe that it does not exhibit any squeezing in 
polarization as per the general criteria \cite{luis-korolkova,prakash-shukla} for a general component of Stokes operator vector. 

\section{Polarization Squeezing in the mixed beam through Linear Beam splitter}
After mixing of the coherent beam and squeezed vacuum, the mean values and variances of the Stokes operators at the output port
$\bm 3$ represented by Eq.~(\ref{eq9}) for $\alpha_{x,y}=|\alpha_{x,y}|e^{i\phi_{x,y}}$ can be calculated as

\begin{eqnarray}\label{eq16}
\expec{S_0}&=&t_{x}^2|\alpha_x|^2+t_{y}^2|\alpha_y|^2+(r_{x}^2-r_{y}^2)s^2,\nonumber\\
\expec{S_1}&=&t_{x}^2|\alpha_x|^2-t_{y}^2|\alpha_y|^2+(r_{x}^2+r_{y}^2)s^2,\nonumber\\
\expec{S_2}&=&2t_x t_y |\alpha_{x}||\alpha_y|\cos (\phi_y-\phi_x),\nonumber\\
\expec{S_3}&=&2t_x t_y |\alpha_{x}||\alpha_y|\sin (\phi_y-\phi_x), 
\end{eqnarray}

and

\begin{eqnarray}
\label{eq17}
V_1&=&2s^2 r_{x}^2 t_{x}^2|\alpha_x|^2+2s^2 r_{y}^2t_{y}^2|\alpha_y|^2+t_{x}^2|\alpha_x|^2+t_{y}^2|\alpha_y|^2\nonumber\\
&&-4csr_x r_y t_x t_y|\alpha_x||\alpha_y|\sin(\phi_x+\phi_y)\nonumber\\
&&+(r_{x}^2+r_{y}^2)s^2-(r_{x}^2-r_{y}^2)s^4,\nonumber\\
V_2&=&2s^2 t_{x}^2 r_{y}^2|\alpha_x|^2+2s^2 r_{x}^2t_{y}^2|\alpha_y|^2+t_{x}^2|\alpha_x|^2+t_{y}^2|\alpha_y|^2\nonumber\\
&&-4csr_x r_y t_x t_y|\alpha_x||\alpha_y|\sin(\phi_x+\phi_y)\nonumber\\
&&+4 r_{x}^2 r_{y}^2 s^4+2 r_{x}^2 r_{y}^2(2s^4+s^2),\nonumber\\
V_3&=&2s^2 t_{x}^2 r_{y}^2|\alpha_x|^2+2s^2 r_{x}^2t_{y}^2|\alpha_y|^2+t_{x}^2|\alpha_x|^2+t_{y}^2|\alpha_y|^2\nonumber\\
&&-4csr_x r_y t_x t_y|\alpha_x||\alpha_y|\sin(\phi_x+\phi_y)\nonumber\\
&&-4 r_{x}^2 r_{y}^2 s^4+2 r_{x}^2 r_{y}^2(2s^4+s^2),\nonumber\\
\end{eqnarray}
We now consider the polarization squeezing under the approximation $t_x|\alpha_x|=t_y|\alpha_y|=A>>c$ , {\it i.e.}, the transmitted parts of coherent light in 
the $x$ and $y$ modes have the same amplitude which is very large as compared to $\cosh kt$. This is to note that, if we do not consider large interaction 
times which allows us to ignore the higher order terms in $s$, {\it i.e.} $s^2$ and $s^4$, we can test the polarization squeezing 
along all the three Stokes operators as shown below. For the Stokes operator $\hat S_1$, we have
\begin{eqnarray}
&&V_1=2s^2(r_{x}^2+r_{y}^2)A^2-4cs~r_x r_y A^2 \sin (\phi_x+\phi_y)+2A^2,\nonumber\\
&&{\expec{\hat S_2}}^2+{\expec{\hat S_3}}^2=8A^4.\nonumber
\end{eqnarray} The squeezing factor for $\hat S_1$ obtained by plugging in these values in Eq.~(\ref{eq6}) is
\begin{equation}
\label{eq18}
\mathcal{S}_{1}=\frac{1}{\sqrt{2}} [1+s^2(r_{x}^2+r_{y}^2)-2cs r_x r_y \sin (\phi_x+\phi_y)].
\end{equation}
This expression has the minimum value for $\phi_x+\phi_y=\pi/2$ and the maximum
polarization squeezing would be obtained for $\tanh2kt={2 r_x r_y}/{r_{x}^2+r_{y}^2}$.
Therefore, the maximum polarization squeezing quantified by the minimum value of polarization squeezing factor is
\begin{equation}
\label{eq19}
\mathcal{S}_{1 min}=\frac{1}{\sqrt{2}} \bigg[1-{[min(r_x,r_y)]}^2\bigg], 
\end{equation}
where $ min(r_x,r_y)$ is $r_x$ if $r_x<r_y$,~~$r_y$ if $r_y<r_x$ and $r_y=r_x$ 
if both are equal.\\

For second Stokes operator $\hat S_2$ as per the same criterion in Eq.~(\ref{eq6}), squeezing factor can be written as
\begin{equation}
\label{eq20}
\mathcal{S}_{2}=\frac{1+s^2(r_{x}^2+r_{y}^2)-2cs r_x r_y 
\sin (\phi_x+\phi_y)}{\sin (\phi_y-\phi_x)}. 
\end{equation}
This expression can be minimized  for $(\phi_x+\phi_y)=\pi/2$ and 
$(\phi_y-\phi_x)={\pi}/2$, and it leads to maximum squeezing for $\tanh kt={2 r_x r_y}/{r_{x}^2+r_{y}^2}$ with squeezing factor 
resulting into the same expression as in the previous case. It is given by
\begin{equation}
\label{eq21}
\mathcal{S}_{2 min}=1-{[min(r_x,r_y)]}^2.
\end{equation}
In a similar way, we obtain the minimum squeezing factor in case of Stokes operator
$\hat S_3$ for $(\phi_x+\phi_y)=\pi/2$ and $(\phi_y-\phi_x)=0$ with a condition $\tanh kt={2 r_x r_y}/{r_{x}^2+r_{y}^2}$, as
\begin{equation}
\label{eq22}
\mathcal{S}_{3 min}=1-{[min(r_x,r_y)]}^2.
\end{equation}
The squeezing factor for all the three components of Stokes operator vector can therefore be written as
\begin{equation}
\label{eq23}
\mathcal{S}_{1 min}=\frac{1}{\sqrt{2}}[1-r^2],\mathcal{S}_{2 min}= \mathcal{S}_{3 min}=1-r^2,
\end{equation}
where ~$r=min(r_x,r_y)$.\\

\begin{figure}[bt]
\centering
\includegraphics[width=\linewidth]{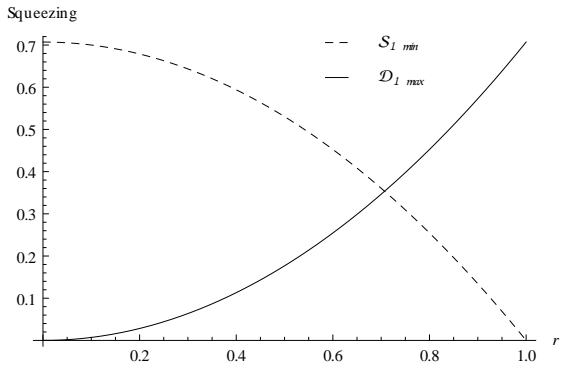}
\caption{Variation of $\mathcal{S}_{1min}$ and $\mathcal{D}_{1max}$ with $r$ showing maximum polarization squeezing.}
\label{f2}
\end{figure}

Above results show that the squeezed vacuum is not polarization squeezed in itself, 
but when mixed with a coherent beam, the output beam exhibits polarization squeezing along 
all the three components of Stokes operator vector for $(\phi_x+\phi_y)=\pi/2$
but different combinations of $(\phi_x,\phi_y)$. We observe that the output light is most squeezed in polarization along the 
Stokes parameters $\hat S_1$ and equally squeezed along $\hat S_2$ and $\hat S_3$. The variation of squeezing factor and 
degree of squeezing with $r$, corresponding to maximum polarization squeezing in $\hat{S}_2$ and ${\hat S}_3$ is shown in Fig.~\ref{f2}.

\section*{Discussion of Result}
Looking at the above expressions for squeezing factor, one can observe that, it can be
made less than one by choosing $r$ small. This gives high degree of polarization squeezing. 
As a special case, if we consider this linear beam splitter to be symmetric one, 
$ r_x=r_y=\frac{1}{\sqrt{2}}$ that gives the minimum squeezing factor $\mathcal{S}_{1 min}=0.35$,
{\it i.e.}, $\mathcal{D}_{1 min}=0.65$. However, in the case of $\hat S_2$ and $\hat S_2$, we have $\mathcal{S}_{2 min}=\mathcal{S}_{2 min}=0.35$, 
and  $\mathcal{D}_{2 max}=\mathcal D_{3 min}=0.50$. This reveals a maximum of $ 65 \% $ squeezing at the output port $\bm 3$ and this is observed along $\hat S_1$. 
We therefore observe that squeezed vacuum on mixing with a coherent radiation through a beam splitter leads 
to polarization squeezing in the output light. This is important because the nature of non-classicality 
changes during the interaction and up to $65\%$ squeezing in polarization is obtained without initial non-classicality of same nature. 
In our next manuscript, we are further exploring the simultaneous squeezing of orthogonal components of Stokes operator vector.

\section*{Acknowledgements}
We would like to thank Hari Prakash for his interest and critical comments.
\end{document}